# Model uncertainty in accelerator application simulations[1,2]

Vitaly Pronskikh[3]

Fermi National Accelerator Laboratory, Batavia, IL 60510-5011, USA

Abstract

Monte-Carlo nuclear reaction and transport codes are widely used to devise accelerator-based nuclear physics experiments; at the same time, many experiments are performed to validate the Monte-Carlo codes, which can be used for the design of full-scale nuclear power applications or the design of new benchmark experiments. Dedicated model benchmark studies investigate a broad range of nuclear reactions and quantities. Examples of these include isotope formation or secondary particle fluxes that result from the interactions of GeV-range hadrons with monoisotopic targets, which can be used to assess the respective systematic uncertainty of models. Such benchmark studies, as well as many nuclear application experiments and simulations carried out by various groups over the last few decades, enable us to draw methodological lessons. In this work, model uncertainty determined based on available experimental data allow us to identify the effects of practitioner expertise as well as the design of codes (user access to micro-scale parameters) on the range of uncertainties. We found that in cases when simulations are performed by code developers or users that are very experienced in performing simulations, the model to experiment quantity ratios generally agree with the limits determined by dedicated benchmark studies. In other cases, the ratios generally tend to be either smaller (underestimation of model error) or larger (overestimation of model error). A plausible explanation of the aforementioned effects is suggested.

# Introduction

A large number of nuclear applications, including research on advanced nuclear reactor concepts [1, 2] and activation studies for the shielding design of high-energy physics experiments [3, 4], not only require elaborated experimental techniques and accelerator facilities but also demand nuclear model codes that are capable of supporting experimental activities. In as much as the design of future research facilities and power stations is grounded in an extensive application of Monte-Carlo computer simulation codes, validation of these codes becomes the most important task for nuclear physics experiments. Many studies have been performed to validate computer models of nuclear reactions incorporated in modern codes; these studies determine model uncertainty for various projectiles and targets, especially for the spallation domain [5; 6; 7; 8; 9; 10; 11].

Despite the fact that most of the models used in simulations have been benchmarked in broad ranges of parameters, experimental groups report comparisons of calculated to experimental ratios that suggest that the model to experiment agreement is not always consistent with dedicated benchmark studies. As most of the codes are multiscale, i.e., they are comprised of a level of intranuclear processes and level of macroscopic processes, recent research suggests that in cases when model users are able to make changes, they can also change micro-scale parameters while working at the level of macroscopic model construction (dealing with objects like targets, shielding, etc.). Such adjustable micro-scale parameters can be those of nuclear cascades to perform so-called piecemeal adjustments of the model; the entanglement of verification and validation of model parameters can occur resulting in underestimation of the model error (systematic uncertainty) [12; 13]. The aim of this work is to define micro-scale parameters for the most popular Monte Carlo codes (such as FLUKA [15], Geant4 [14], MARS15 [17], MCNPX [18], PHITS [16]) and their accessibility by users as well as to determine correlations between developer and non-developer applications of codes and model uncertainty resulting from such use.

## Classifications of codes based on access to low-level parameters

Table I presents a list of major micro-scale model parameters for Monte Carlo codes most frequently used in accelerator-based nuclear applications as well as designate the accessibility of such parameters by users, usually, through their input decks. In addition to the model validity energy ranges, which parameters more or less clearly follow from model documentation and physics considerations (these parameters are user-accessible in all codes considered), there are a number of parameters that are validated in separate micro-scale studies and are not always made adjustable by users. The latter parameters can serve, for instance, intranuclear cascade and evaporation parameters (including nuclear densities, particular mechanisms, etc.), and reaction cross sections. Detailed descriptions of the micro-scale physics models available in manuals (with comprehensible comments) belong to such a level.

With regards to the access to certain model parameters, we ascribe the codes to two types, A and B:

1. Codes with minimal access to micro-scale parameters (A): FLUKA, MARS15, PHITS
2. Codes with a full access to micro-scale parameters (B): GEANT4, MCNPX/MCNP6

## Classification of codes based on user expertise

In this work, we also propose to distinguish three groups of code usage depending not only on the access to micro-scale parameters described in the previous paragraph but also on whether the code is used by a developer or a regular user:

1. Group I. Type A or B codes used by developers
2. Group II. Type A codes used by non-developers
3. Group III. Type B codes used by non-developers

The justification for such a division is given as follows: an expert (developer's) deployment of a code of either A or B type can be expected to be equally competent, whereas for the case of non-developers, the less parameters are amenable to adjustment by the user, the less the model error generated by incomplete understanding of the underlying physics processes. In that respect, Group I code use can be deemed least error-prone, while Group II and, especially, Group III code use can be anticipated to exhibit an increase in model uncertainty due to a potentially imperfect choice of parameters.

Table I. User access to micro-scale parameters in Monte-Carlo codes

| Code / User access | PHITS | FLUKA | MARS15 | MCNPX | GEANT4 |
|---|---|---|---|---|---|
| Total reaction cross sections | + | - | - | - | + |
| Angular distribution of secondaries in cascade | + | - | - | - | + |
| Adding/excluding particular mechanisms for particle interactions and decays | + | + | + | + | + |
| Choice of models (evaporation, pre- | + | + | + | + | + |
| equilibrium and/or multifragmentation) | | | | | |
| Choice of model energy ranges | + | + | + | + | + |
| Access to the source code of the physics models (e.g., intranuclear cascades) or their key constants | - | - | - | + | + |
| Availability of the description of the formalization of the physics models in the manual | - | - | - | + | + |

Based on the uncertainty classification, we propose to ascribe each result group that takes into consideration whether the code allows immediate access to its micro-scale parameters as well as whether the users include developers of the particular code. Let us first discuss the particular parameters and in what way they can be related to a certain scale.

## Micro- and macro-scales for nuclear simulations

The nuclear interaction and transport codes rely on two different scales, which we call the micro-scale and macro-scale level. The micro-scale is a level where the code simulates the intranuclear processes, which occur when an individual energetic particle or nucleus interacts with another nucleus. The typical simulated stages of that process are given as follows: First, intranuclear cascade (INC), in which nucleons follow straight trajectories, and, similar to a billiard ball, knock out nucleons. There are slight variations between codes and models in their approach to INC; for example, the stopping criterion (energy threshold or exceeding allowed interaction time), emission of clusters or the addition of a pre-equilibrium stage [19]. CEM as well as LAQGSM [19] and INCL [20] are built-in INC generators in MCNPX, while the combination of the former two forms the baseline approach in MARS15. There are alternatives to INC that take into account non-binary interactions; for example, Quantum-Molecular Dynamics models implemented in PHITS [21] and FLUKA. Concerning the option to allow users to vary INC parameters, there are two main approaches. Some codes, such as [15], [16], [17] follow the approach that was expressed by the authors of a benchmark activity as "[t]he consensus was that

[the options to fix the parameters on a physics basis with the risk of poorer agreement with some experimental data] should be favoured to have a general good predictive power." [22]. Other codes, for example [14] and [18] allow the user of macro-models access to more parameters of both scales, including the micro-scale. The micro-scale simulation in the codes requires such parameters as nuclear medium parameters and corrections, nuclear potential $V_N$ and shape, Pauli blocking, range and mechanisms (like coalescence) of composite particle emission, criterion for transition between INC and pre-equilibrium, multiplicities of emitted particles etc. to be either chosen (Group B) or taken as a default preset by developers (Group A). The total reaction cross section for normalization is also among the important parameters.

Second, the after-cascade deexcitation of a nucleus is usually described as a competition between evaporation, fission, and multifragmentation. The models at this stage are used to specify either emission of nucleons or nuclei from excited nuclei [23, 24], its symmetric or asymmetric fission (both intermediate mass and light mass), for example, using the Bohr-Wheeler formalism. Additionally, this stage (belonging also to the macro-scale) can include the emission of light fragments using a multifragmentation model [25], implemented in PHITS. The key parameters at this stage are allowed deexcitation channels and their combinations, fission barriers, Coulomb barriers, fission and evaporation level densities, multiplicities of emitted particles, etc. Recoil and residual energies, masses, charge, momenta can also be regarded as parameters of importance.

INC and deexcitation make up the microscopic scale for simulations of nuclear systems. The macro-scale is that at which code users create models of macroscopic objects such as targets, moderators, shielding blocks etc. On the one hand, such objects consist of many nuclei in which micro-scale processes occur (like energy deposition or radiation damage). On the other hand, the macro-scale processes involve a multitude of intranuclear cascades and interactions, where particles and fragments resulting from individual INC, evaporations, and fissions propagate to an ample number of other nuclei in the system (e.g., accelerator targets) and induce microscopic processes in them. The macroscopic systems require completely different sets of parameters to deploy models, such as macroscopic volumes, densities, object shapes, distances, materials, distributions of electric and magnetic fields, accelerated beam shapes, etc. Concerning the scope of expertise [13], end users of the simulation codes who are not developers are expected to be expert primarily in the macroscopic applications and their properties described by the macro-scale parameters. However, those are not expected to be equally proficient in the underlying microscopic models and parameters except for a general understanding of the physics concepts the models are premised upon.

## Model uncertainty determined by developers

The benchmark [22] results can be used for the assessment of uncertainties for nucleon-induced reactions in the energy range between 20 MeV and 3 GeV on targets ranging from carbon to uranium. The comparison team [22] made several conclusions concerning the agreement between data and simulation results [5]. They found out that for neutron multiplicities, depending on the mass number of the target nucleus, the agreement is between 50% and a factor of 2; for differential neutron spectra in the energy range 1 MeV to 1 GeV, the level of agreement is generally between 20% and 70%. For most of the models used in the comparison, the agreement in the residual nuclei is generally between 50% and a factor of 3 (for Z > 10). Another series of benchmark experiments

[7] has determined that for a set of medium-mass and heavy nuclei used as targets for 660-MeV protons, the theoretical to experimental residual production cross section ratios vary between 1.3 to 3.3 and 3.2 to 7.3; however, because of the large uncertainty due to averaging over more than a hundred residuals, this result only allows one to rule out the model combinations that are less adequate for describing spallation reactions on particular isotopes. The most promising models used in [7] are also those deployed in the [5] comparison; these models (like CEM or LAQGSM) are also implemented in [17] and [18]. The good agreement between neutron multiplicities and residual nuclides suggests that the results from the comparison [5] are consistent. However, we have to mention here that the calculations for the comparisons presented in [5] were conducted by code developers, which explains not only the consistency but also the optimal choice of the micro-scale model parameters used in simulations. In what follows, we assume that a 50% to 300% model uncertainty is attainable by an expert user and undertake an analysis of the discrepancies between the developer and non-developer simulations for a number of studies that report measurements of both micro-scale processes (nuclear reactions on thin targets) and macro-scale processes (activation reactions in samples by neutrons emitted from extended irradiated targets).

## Discussion

Table II presents calculated to experimental ration for several key quantities critical for nuclear physics experiments employing nuclear reaction models. The data and calculations were obtained and performed by various groups using different techniques and codes over two decades of studies for 1) interactions of 600 MeV to 6 GeV protons and deuterons with extended (500-kg) natural uranium targets to determine residual production in activation samples placed in secondary particle fluxes; 2) interactions of 600-MeV protons with thin medium-mass and heavy targets with the aim to validate Monte-Carlo computer codes; 3) interactions of a 120-GeV proton beam with a beam absorber in a neutrino experiment for the determination of the activation of Al samples by secondary particle fluxes. All discussed experimental groups relied on the activation technique and gamma-spectroscopy of activation samples by means of semiconductor detectors, which makes it easier to compare their results because of comparable systematic uncertainties.

Measurements of [6] and [7] were carried out using thin targets, and nine or eleven codes were used to simulate the cross sections for the residual production. Although the calculated to experimental ratios vary from one to seven, a more detailed look allows one to identify the codes that provide the most consistent description of the reaction studies; among these codes are the CEM and LAQGSM codes that are incorporated in MCNPX and MARS15, as discussed above. Therefore, the ratios for such experiments can only be used for relative spallation code validation and benchmarks.

The ratios based on the developers' use of the codes that we marked as Group I, range between one and three, and, hence, are in a remarkable agreement with the benchmark reported by [5]. As for the Groups II and III, their ratios fall into two categories: those in which they are significantly greater than 3 (e.g., 5 or 6) and those smaller than 3 (2 or less). The former can be explained by the presumably "black box" approach used where model parameters were not chosen optimally. The latter is, in our opinion, due to the possible piecemeal adjustments made between the model and data caused by tuning the model parameters and their optimization for a particular data set (exceedingly good agreement with data is observed for Group III); e.g., non-developer use of the codes allowing access to both micro and macro-level parameters. The former pitfall is more typical for an unexperienced code user, whereas the latter is more frequently experienced by more skilled users who are, nevertheless, not acquainted with the entire structure and features of the computational model. There is, however, an example of [11], who demonstrate ratios in the same range as presented by developers as well as those in agreement with the benchmark. We can explain this as being due to the substantial experience of that particular group in computer simulations (even if they are not developers of the codes used in their paper).

Table II. Calculated to experimental ratios determined from the results from various groups.

| Paper | Beam/Target | Sample | Reaction or quantity | Code | Group | Ranges of averaged ratios $\sigma_{calc}/\sigma_{exp}$ |
|---|---|---|---|---|---|---|
| [6] | p(660 MeV)+$^{129}$I | $^{129}$I | 22 spallation products | 11 codes | I | 1.3(12)-3.3(28) |
| [7] | p(660 MeV)+$^{237}$Np | $^{237}$Np | 32 spallation products | 9 codes | I | 3.6(19)-7.3(45) |
| [7] | p(660 MeV)+$^{241}$Am | $^{241}$Am | 37 spallation residuals | 9 codes | I | 3.2(23)-7.1(43) |
| [8] | d(1-4 AGeV)+500kg $^{238}$U | $^{237}$Np $^{238}$Pu $^{239}$Pu | Neutron induced fission | MCNPX | III | 1.25(15)-0.91(8) |
| | | | | MARS15 | I | 1.43(20)-0.71(5) |
| [26] | P(660 MeV)+500kg $^{238}$U | $^{59}$Co | (n,γ); (n,xn) | MCNPX | III | 0.78(8)-1.27(13) |
| | | | | MARS15 | I | 0.36(4)-1.50(15) |
| [27] | P(660 MeV)+500kg $^{238}$U | $^{127}$I | (n,γ); (n,xn) | Geant4 BERT_HP | III | 0.177(8)-5.6(3) |
| | | | | Geant4 BIC_HP | III | 0.177(8)-3.13(17) |
| | | | | Geant4 INCLXX_HP | III | 0.189(8)-2.44(13) |
| [28] | d(6 GeV)+500kg $^{238}$U | $^{232}$Th | (n,γ); fission | FLUKA (RQMD) | II | 1.64(5)-6.6 |
| [11] | d(4 GeV)+500kg $^{238}$U | $^{238}$U; $^{232}$Th | fission | MCNPX (INCL4/CEM03) | III | 1.00(12)-2.44(18) |

| [9] | P(120 GeV) + 10 cm thick W target | 15° | Neutron flux | FLUKA PHITS | II I | 0.275-2.08 0.275-2.71 |
| | P(120 GeV)+94 cm long graphite target | Integrated circuits dosimeter (IC) | Ambient dose equivalent rates | MARS15 | I | 1-2.65 |
| [10] | P(120 GeV)+ 244 cm Al+ 231 cm Iron+ 91.4 cm concrete | Al, Au | Residuals | PHITS+MARS15 | I | 0.32-2.2[4] |

Figure 1 allows for a visual comparison of the data from Table II of 300 % and 30 % systematic uncertainty levels. Each position on the horizontal axis of the figure for a particular group corresponds to a particular paper, while the data for different groups at the same position on that axis are independent. The points are connected by eye-guide lines. Red horizontal lines represent 300% deviation from unity, whereas brown lines represent 30% deviation. (Based on the discussion above as well as benchmark experiments, we suppose that uncertainties smaller than 30% are currently unattainable provided the model parameter set is universal.) Data presented in Figure 1 allow us to make the following observations:

1. For the 300% uncertainty range, all Group I data are either within that range or (in one case) close to it, whereas in several cases Group II and Group III points lie far beyond their limits. This points out that non-developer simulations in many cases either significantly overestimate or underestimate measurement.
2. For the 30% uncertainty range, based on the data taken into consideration, the Group III data often lie within it, with one exception of a Group I Min data point, which also falls within those limits. Given that the Group III simulations employ type B codes, where a nonexpert user has access to micro-scale parameters, as described above—for example, total reaction cross sections or tunable parameters of the intranuclear cascade generators—one might assume that excessively small calculated-to-experimental ratios can be assigned to the outcome of a piecemeal adjustment of those parameters to the particular data set. If our assumption is true, such an underestimation of uncertainty has such a drawback that the chosen parameter set might lack universality if applied to other independently obtained experimental data sets. In the case of the Group I data point that has a Min I value (lowest margin) close to a unity, we explain it by assuming that the set of parameters used might fall occasionally close to experimental values, which is supported by the corresponding Max I data point (higher margin) that is greater than 200%. Rather than for Group I, in two cases of the Group III data, both the Min and Max are within the 30 % limit, which seems overly optimistic.

---

[4] [10] mention that one particular residual, namely, $^{185}$Os, has a calculation to experimental ratio of close to 5; however, insufficient production of this isotope by the models employed is elucidated in their paper taking into account the multifragmentation mechanism of its formation, which was not implemented in the code versions used.

3. The plausible evidence in favor of the presupposition that nonexpert usage of type A codes (Group II) is more adverse than that of type B ones (Group III) might serve the observation that in Group II simulations, in which there is no user access to micro-level parameters, calculations overestimate only experimental data, whereas in the case of Group III, calculations often underestimate measured values. In our opinion, underestimating uncertainties is less benign than overestimating them (although we concede that the correctness of such a statement is conditional within the scope of the code application).
4. Also important is that Group II uncertainties are greater than 30% in the studied cases; although, in one case, both Min II and Max II calculations overestimate experimental numbers (in all other cases, Min points are below unity, and the Max one is above it), which suggests an imperfect choice of simulation models or their parameters.

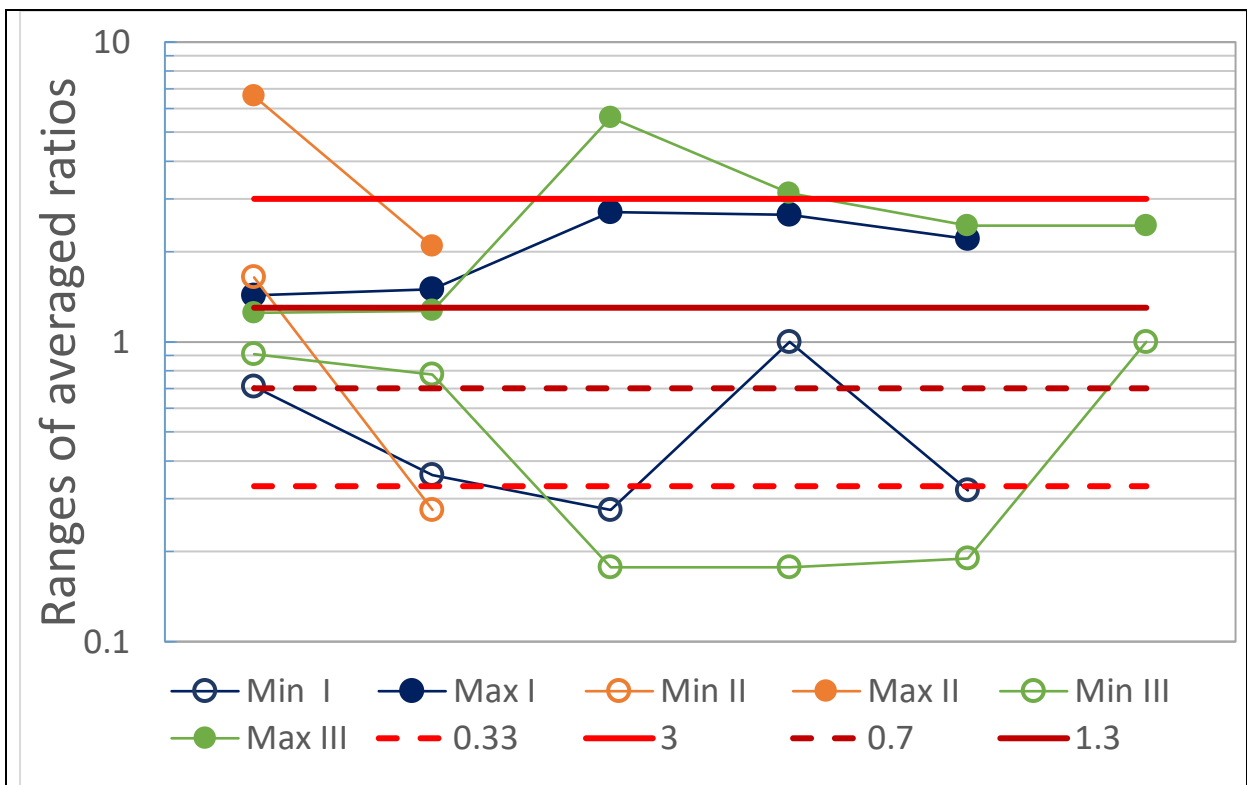

Figure 1. Ranges of averaged ratios of calculated-to-experimental quantities. Min: lowest ratio of the range in the particular paper. Max: highest ratio of the range in the particular paper. I: Codes of either A or B type used by developers. II: codes of A type used by non-developers. III: codes of B type used by non-developers.

## Conclusion

In this study, the model systematic errors arising from computer model simulations of nuclear reaction-related quantities in a number of macro-scale accelerator-based nuclear applications were assessed to indirectly evaluate the effect of users' expertise on the consistency of model errors, as well as the accessibility of the micro-scale parameters in the most popular Monte Carlo codes. The model quantities expressed in terms of calculated to experimental quantity ratios were comparatively analyzed with those obtained from benchmark experiments and calculations. Five Monte Carlo codes, as well as their applications to several nuclear accelerator experiments, were analyzed. The codes and modes of use were classified into two and three categories depending on the users' expertise as well as adjustability of their macro-scale parameters by external users, respectively.

The study indicates that the simulation results produced by model developers (and, in certain cases, by especially experienced users) usually fall within a factor of 2-3 of the experimental values, which is consistent with the dedicated benchmarks. On the other hand, the calculated to experimental ratios produced by other users are often either below that range (underestimate model error due to over-optimization) or above it (overestimate model error). The former can be understood if one takes into account the possibility of piecemeal adjustment of micro-scale parameters, while the latter may point to a non-optimal choice of controlling values and working regimes for the model code.

Such an outcome of our study is consistent with a body of methodological research suggesting that inseparability of micro- and macro-scales in simulations entails their entanglement, and, hence, increased systematic uncertainty for such simulations. The paper presents the author's analysis of the methodological approaches used by code users, assessing their weaknesses and advantages, and providing recommendations for the elaboration and development of those methods. One possible improvement of practice that follows from our analysis is developing extensive documentation for codes and having code developers provide user tutorials. In our paper, we elicit that developers and very experienced users achieve more consistent results using codes. The paper also provides a classification of codes based on their internal structure and explains which codes are safer for less-experienced developers to use, as well as why, and to what extent those codes are safer. Our analysis allows estimation of an approximate anticipated systematic error when certain users apply certain codes.

## Acknowledgements

The author is indebted to two anonymous referees for their careful reading of the manuscript and many useful comments that helped improve it.

This manuscript has been authored by Fermi Research Alliance, LLC under Contract No. DE-AC02-07CH11359 with the U.S. Department of Energy, Office of Science, Office of High Energy Physics.

## References

1. Charles D. Bowman, Accelerator-driven systems for nuclear waste transmutation, Annual Review of Nuclear and Particle Science 1998 48:1, 505-556.
2. C. Rubbia, S. Buono, Y. Kadi and J.A. Rubio, Fast neutron incineration in the energy amplifier as alternative to geologic storage: the case of Spain, CERN/LHC/97-01(EET).

3. Di Benedetto, V., Gatto, C., Mazzacane, A., Mokhov, N. V., Striganov, S. I., Terentiev, N. K., A study of muon collider background rejection criteria in silicon vertex and tracker detectors, Journal of Instrumentation, Volume 13, Issue 09, pp. P09004 (2018).
4. V. S. Pronskikh. Radiation Studies for the Mu2e Experiment: A Review. Modern Physics Letters A, Vol. 28, No. 19, 1330014 (2013).
5. Jean-Christophe, David, Detlef Filges, Franz Gallmeier, Mayeen Khandaker, Alexander Konobeyev, Sylvie Leray, Guenter Mank, Alberto Mengoni, Rolf Michel, Naohiko Otuka, and Yair Yariv, Benchmark of Spallation Models, Progress in Nuclear Science and Technology, Vol. 2, pp.942-947 (2011).
6. V. S. Pronskikh, J. Adam, A. R. Balabekyan, V. S. Barashenkov, V. P. Dzhelepov, S. A. Gustov, V. P. Filinova, V. G. Kalinnikov, M. I. Krivopustov, I. V. Mirokhin, A. A. Solnyshkin, V. I. Stegailov, V. M. Tsoupko-Sitnikov, J. Mrazek, R. Brandt, W. Westmeier, R. Odoj, S. G. Mashnik, A. J. Sierk, R. E. Prael, K. K. Gudima, M. I. Baznat, Study of Proton Induced Reactions in a Radioactive $^{129}$I Target at $E_p$ = 660 MeV, Proc. of the International Workshop on Nuclear Data for the Transmutation of Nuclear Waste (TRAMU@GSI), GSI-Darmstadt, Germany, September 1-5, 2003; LA-UR-04-2139.
7. S. G. Mashnik, V. S. Pronskikh, J. Adam, A. Balabekyan, V. S. Barashenkov, V. P. Filinova, A. A. Solnyshkin, V. M. Tsoupko-Sitnikov, R. Brandt, R. Odoj, A. J. Sierk, R. E. Prael, K. K. Gudima, M. I. Baznat, Analysis of the JINR p (660 MeV) I-129, Np-237, and Am-241 Measurements with Eleven Different Models, Proc. Seventh Specialists' Meeting on Shielding Aspect of Accelerators, Targets and Irradiation Facilities, SATIF-7, Sacavem (Lisbon), Portugal, May 17-18, 2004, (NEA/OECD, Paris, 2005), pp. 231-241; E-print: nucl-th/0407097; LA-UR-04-4929.
8. L. Zavorka, J. Adam, A.A. Baldin, P. Caloun, V.V. Chilap W.I. Furman, M.G. Kadykov, J. Khushvaktov, V.S. Pronskikh, A.A. Solnyshkin, V. Sotnikov, V.I. Stegailov, M. Suchopar at al., Neutron-induced transmutation reactions in 237Np, 238Pu, and 239Pu at the massive natural uranium spallation target, Nuclear Instruments and Methods in Physics Research, 349, 31-38, 2015.
9. Hiroshi Nakashima, Nikolai V. Mokhov, Yoshimi Kasugai, Norihiro Matsuda, Yosuke Iwamoto, Yukio Sakamoto, David Boehnlein, Anthony Leveling, Kamran Vaziri, Richard Coleman, Douglas Jensen, Erik Ramberg, Aria Soha, Toshiya Sanami, Hiroshi Matsumura, Masayuki Hagiwara, Akihiro Toyoda, Hiroshi Iwase, Hideo Hirayama, Takashi Nakamura, Shun Sekimoto, Hiroshi Yashima, Tsuyoshi Kajimoto, Nobuhiro Shigyo, Kenji Ishibashi, Noriaki Nakao, Norikazu Kinoshita, Koji Oishi, Hee-Seock Lee and Koji Niita, Research activities on JASMIN: Japanese and American Study of Muon Interaction and Neutron detection, Progress in Nuclear Science and Technology Volume 4 (2014) pp. 191-196.
10. Norihiro Matsuda, Yoshimi Kasugai, Hiroshi Matsumura, Hiroshi Iwase, Akihiro Toyoda, Hiroshi Yashima, Shun Sekimoto, Koji Oishi, Yukio Sakamoto, Hiroshi Nakashima, Takashi Nakamura, David Boehnlein, Gary Lauten, Anthony Leveling, Nikolai Mokhov, Kamran Vaziri, Activation detector measurements at the hadron absorber of the NuMI neutrino beamline at Fermilab, Progress in Nuclear Science and Technology Volume 4 (2014) pp. 337-340.
11. S.R. Hashemi-Nezhad, N.L. Asquith, V.A. Voronko, V.V. Sotnikov, Alina Zhadan, I.V. Zhuk, A. Potapenko, Krystsina Husak, V. Chilap, J. Adam, A. Baldin, A. Berlev, W. Furman, M. Kadykov, J. Khushvaktov, I. Kudashkin, I. Mar'in, M. Paraipan, V. Pronskikh, A. Solnyshkin, S. Tyutyunnikov, Transmutation of uranium and thorium in the particle field of the Quinta sub-critical assembly, Nucl. Instr. Meth. in Phys. Res. Section A: Accelerators, Spectrometers, Detectors and Associated Equipment, 883 (2018) 96-114.
12. Winsberg, E. (2010). Science in the age of computer simulations. Chicago, London: The University of Chicago Press.

13. Pronskikh V. Computer modeling and simulation: Increasing reliability by disentangling verification and validation, Minds and Machines. 2019. Vol. 29. P. 169–186.
14. Agostinelli S. et al. Geant4—a simulation toolkit, Nuclear Instruments and Methods in Physics Research Section A: Accelerators, Spectrometers, Detectors and Associated Equipment, Volume 506, Issue 3, 2003, Pages 250-303.
15. T.T. Böhlen, F. Cerutti, M.P.W. Chin, A. Fassò, A. Ferrari, P.G. Ortega, A. Mairani, P.R. Sala, G. Smirnov and V. Vlachoudis, The FLUKA Code: Developments and Challenges for High Energy and Medical Applications, Nuclear Data Sheets 120, 211-214 (2014).
16. T. Sato, K. Niita, N. Matsuda, S. Hashimoto, Y. Iwamoto, S. Noda, T. Ogawa, H. Iwase, H. Nakashima, T. Fukahori, K. Okumura, T. Kai, S. Chiba, T. Furuta and L. Sihver, Particle and Heavy Ion Transport Code System PHITS, Version 2.52, J. Nucl. Sci. Technol. 50, 913-923 (2013).
17. N.V. Mokhov and C.C. James, "The MARS code System User's Guide, Version 15 (2016)", Fermilab-FN-1058-APC (2017); https://mars.fnal.gov; Mokhov, N., Aarnio, P., Eidelman, Yu, Gudima, K., Konobeev, A., Pronskikh, V., Rakhno, I., Striganov, S., Tropin, I., 2014. MARS15 code developments driven by the intensity frontier needs. Prog. Nucl. Sci. Technol. 4, 496–501, FERMILAB- CONF-12-635-APC.
18. Denise B. Pelowitz et al., "MCNPX 2.7.0 Extensions", LA-UR-11-02295 (2011).
19. S.G. Mashnik, K.K. Gudima, R.E. Prael, A.J. Sierk, M.I. Baznat, N.V. Mokhov, CEM03.03 and LAQGSM03.03 Event Generators for the MCNP6, MCNPX, and MARS15 Transport Codes, Invited Lectures Presented at the Joint ICTP-IAEA Advanced Workshop on Model Codes for Spallation Reactions, February 4–8, 2008 ICTP, Trieste, Italy, Los Alamos National Laboratory report LA-UR-08-2931, 2008 E- print: arXiv:0805.0751v2 [nucl-th]; IAEA Report INDC(NDS)-0530, Vienna, Austria, August 2008, p. 53.
20. A. Boudard, J. Cugnon, S. Leray, and C. Volant. Intra-nuclear cascade model for a comprehensive description of spallation reaction data, Phys. Rev., C66, 044615 (2002).
21. Koji Niita, Satoshi Chiba, Toshiki Maruyama, Tomoyuki Maruyama, Hiroshi Takada, Tokio Fukahori, Yasuaki Nakahara, and Akira Iwamoto, Phys. Rev. C52, 2620 (1995).
22. D. Filges, S. Leray, Y. Yariv, A. Mengoni, A. Stanculescu, G. Mank, Joint ICTP-IAEA Advanced Workshop on Model Codes for Spallation Reactions, International Centre for Theoretical Physics Trieste, Italy  4 – 8 February 2008, IAEA INDC(NDS)-0530 Distr. SC.
23. Furihata S. (2001) The GEM Code - the Generalized Evaporation Model and the Fission Model. In: Kling A., Baräo F.J.C., Nakagawa M., Távora L., Vaz P. (eds) Advanced Monte Carlo for Radiation Physics, Particle Transport Simulation and Applications. Springer, Berlin, Heidelberg
24. J. J. Gaimard et al., "A re-examination of the abrasion-ablation model for the description of the nuclear fragmentation reaction," Nuc. Phys., A531, 709 (1991).
25. J.P. Bondorf, A.S. Botvina, A.S. Iljinov, I.N. Mishustin, and K. Sneppen, "Statistical Multi-fragmentation of Nuclei", Physics Reports, 257, 133-221 (1995).
26. J. Adam, A.A. Baldin, M. Baznat, A.I. Berlev, K.V. Gusak, I.V. Kudashkin, J. Khushvaktov, M. Paraipan, V.S. Pronskikh, A.A. Solnyshkin, V. Sotnikov, V.I. Stegaylov, S.I. Tyutyunikov, V. Voronko, M. Zeman, I. Zhuk, Secondary particle distributions in an extended uranium target under irradiation by proton, deuteron, and carbon beams, Nucl. Inst. Meth. in Phys. Res., A 872 (2017) 87–92.
27. J.H. Khushvaktov, J. Adam, A.A. Baldin, W.I. Furman, S.A. Gustov, Yu.V. Kish, A.A. Solnyshkin, V.I. Stegailov, J. Svoboda, P. Tichy, V.M. Tsoupko-Sitnikov, S.I. Tyutyunnikov, R. Vespalec, J. Vrzalova, V. Wagner, B.S. Yuldashev, L. Zavorka, M. Zeman, Monte Carlo simulations and experimental results on neutron production in the uranium spallation target QUINTA irradiated with 660 MeV protons, Applied Radiation and Isotopes, Volume 137, 2018, p. 102-107.

28. J. Adam, V.V. Chilap, V.I. Furman, M.G. Kadykov, J. Khushvaktov, V.S. Pronskikh, A.A. Solnyshkin, V.I. Stegailov, M. Suchopar, V.M. Tsoupko-Sitnikov, S.I. Tyutyunnikov, J. Vrzalova, V. Wagner, L. Zavorka, Study of secondary neutron interactions with $^{232}$Th, $^{129}$I, and $^{127}$I nuclei with the uranium assembly "QUINTA" at 2, 4, and 8 GeV deuteron beams of the JINR Nuclotron accelerator, Applied Radiation and Isotopes, Volume 107, 2016, pp. 225-233.